\documentclass[11pt]{article}

\usepackage[margin=1in]{geometry}
\usepackage{amsmath,amsfonts,amssymb,amsthm,mathtools,nicefrac,mymath}

\newtheorem{definition}{Definition}

\usepackage{graphicx,xcolor}
\usepackage{caption}
\usepackage{subfig}
\graphicspath{{figures/}}
\usepackage{url}

\usepackage{relsize}

\newcommand{\UR}{U_\mathcal{R}}
\newcommand{\UFRQI}{U_\text{FRQI}}
\newcommand{\UQPIXL}{U_\text{QPIXL}}
\newcommand{\RY}{R_y}
\newcommand{\CX}{\text{CNOT}}

\definecolor{myblue}{rgb}{0,0.4470,0.7410}
\definecolor{myred}{rgb}{0.8500,0.3250,0.0980}
\definecolor{myorange}{rgb}{0.9290,0.6940,0.1250}
\definecolor{mypurple}{rgb}{0.4940,0.1840,0.5560}
\definecolor{mygreen}{rgb}{0.4660,0.6740,0.1880}
\definecolor{mylightblue}{rgb}{0.3010,0.7450,0.9330}
\definecolor{mydarkred}{rgb}{0.6350,0.0780,0.1840}

\usepackage{tikz}
\usepackage{pgfplots}
\usetikzlibrary{calc}
\usetikzlibrary{positioning}
\usetikzlibrary{shapes.multipart}
\usetikzlibrary{pgfplots.groupplots}
\pgfplotsset{
  compat=newest,
  table/header=false,
  tick label style={font=\scriptsize},
  label style={font=\scriptsize},
  legend style={font=\scriptsize},
  legend cell align=left,
  colormap={parula}{
    rgb255=(53,42,135)
    rgb255=(15,92,221)
    rgb255=(18,125,216)
    rgb255=(7,156,207)
    rgb255=(21,177,180)
    rgb255=(89,189,140)
    rgb255=(165,190,107)
    rgb255=(225,185,82)
    rgb255=(252,206,46)
    rgb255=(249,251,14)
  }
}
\pgfplotsset{
  myColOne/.style={myblue},
  myColTwo/.style={myred},
  myColThr/.style={myorange},
  myColFou/.style={mypurple},
  myColFiv/.style={mygreen},
  myColSix/.style={mylightblue},
  myColSev/.style={mydarkred}
}
\pgfplotsset{
  myStyOne/.style={myColOne,thick,mark=+},
  myStyTwo/.style={myColTwo,thick,mark=+},
  myStyThr/.style={myColThr,thick,mark=+},
  myStyFou/.style={myColFou,thick,mark=+},
  myStyFiv/.style={myColFiv,thick,mark=+},
  myStySix/.style={myColSix,thick,mark=+},
  myStySev/.style={myColSev,thick,mark=+},
  myStyRes/.style={black,densely dotted},
}
\pgfplotsset{
  myStyOn/.style={myblue,     thick,mark=asterisk},
  myStyTw/.style={myred,      thick,mark=asterisk},
  myStyTh/.style={myorange,   thick,mark=asterisk},
  myStyFo/.style={mypurple,   thick,mark=asterisk},
  myStyFi/.style={mygreen,    thick,mark=asterisk},
  myStySi/.style={mylightblue,thick,mark=asterisk},
  myStySe/.style={mydarkred,  thick,mark=asterisk},
}
\pgfplotsset{
  every axis/.append style={
    label style={font=\footnotesize},
  }
}
\newcommand{\figname}[1]{\relax}
\newcommand{\fignames}[1]{\relax}
\newcommand{\datfile}[1]{fig/#1.dat}

\usepackage[binary-units]{siunitx}
\usepackage{tabularx,multirow,multicol,booktabs}
\newcolumntype{L}{>{\raggedright\arraybackslash}X}
\newcolumntype{C}{>{\centering\arraybackslash}X}
\newcolumntype{R}{>{\raggedleft\arraybackslash}X}

\usepackage[braket]{qcircuit}
\usepackage{environ}
\newcommand{\myqctmp}[2][0.25]{\Qcircuit @C=#2em @R=#1em @!R}
\NewEnviron{myqcircuit}[1][0.25]{\vcenter{\myqctmp[#1]{0.5} {\BODY}}}
\NewEnviron{myqcircuit*}[2]{\vcenter{\myqctmp[#1]{#2} {\BODY}}}
\newcommand{\controlsq}{*!<0em,.025em>-=-<0em>{\square}} 
\newcommand{\ctrlsq}[1]{\controlsq \qwx[#1] \qw}         

\usepackage[colorlinks,urlcolor=blue,linkcolor=blue,citecolor=blue]{hyperref}
\usepackage[capitalize]{cleveref}

\begin{document}

\title{\bf Quantum pixel representations and compression\\ for $N$-dimensional images}
\author{%
Mercy G. Amankwah\textsuperscript{1,2},
Daan Camps\textsuperscript{1},
E. Wes Bethel\textsuperscript{1},\\
Roel Van Beeumen\textsuperscript{1},
Talita Perciano\textsuperscript{1}
}
\date{\small%
\textsuperscript{1}Computational Sciences Area,
Lawrence Berkeley National Laboratory, CA, United States\\
\textsuperscript{2}Department of Mathematics, Applied Mathematics and Statistics,\\
Case Western Reserve University, OH, United States\\
}

\maketitle

\begin{abstract}
We introduce a novel and uniform framework for quantum pixel representations
that overarches many of the most popular representations proposed in the recent
literature, such as (I)FRQI, (I)NEQR, MCRQI, and (I)NCQI.
The proposed QPIXL framework results in more efficient circuit implementations
and significantly reduces the gate
complexity for all considered quantum pixel representations.
Our method only requires a linear number of gates in terms of the number of pixels and does not use ancilla qubits. 
Furthermore, the circuits only consist of $\RY$ gates and $\CX$ gates making them practical in the NISQ era.
Additionally, we propose a circuit and image compression algorithm that is shown
to be highly effective, being able to reduce the necessary gates 
to prepare an FRQI state for example scientific images by up to 90\% without sacrificing image quality.
Our algorithms are made publicly available as part of \texttt{QPIXL++},
a Quantum Image Pixel Library.
\end{abstract}


\newcommand{\secintro}{Introduction}
\newcommand{\seclitreview}{Related work}
\newcommand{\secqpixl}{QPIXL: Quantum pixel representations}
\newcommand{\secqpixlqc}{QPIXL quantum circuit implementation}
\newcommand{\secfrqi}{FRQI in the QPIXL framework}
\newcommand{\secourfrqi}{Optimal linear gate complexity}
\newcommand{\seccompr}{Compression}
\newcommand{\seccolor}{Other QPIXL mappings}
\newcommand{\secexamples}{Experiments}
\newcommand{\secconclude}{Conclusion}

\newcommand{\refintro}{\Cref{sec:intro}}
\newcommand{\reflitreview}{\Cref{sec:litreview}}
\newcommand{\refqpixl}{\Cref{sec:quantum-image}}
  \newcommand{\refqpixlqc}{\Cref{sec:qpixl-qc}}
\newcommand{\reffrqi}{\Cref{sec:frqi}}
\newcommand{\refourfrqi}{\Cref{sec:ourfrqi}}
\newcommand{\refcompr}{\Cref{sec:compr}}
\newcommand{\refcolor}{\Cref{sec:color}}
\newcommand{\refexamples}{\Cref{sec:examples}}
\newcommand{\refconclude}{\Cref{sec:conclude}}


\section{\secintro}
\label{sec:intro}

The growth in scientific data size and heterogeneity overwhelms current statistical and learning approaches for analysis and understanding.
More specifically, the analysis of image-based data becomes increasingly challenging using current classical algorithms.
Consequently, finding more efficient ways of handling scientific data is an important research priority.

Quantum computing holds the promise of speeding up computations in a wide variety of fields~\cite{Nielsen:2011:QCQI}, including image processing.
One of the research challenges to make quantum computing a viable platform in the post-Moore era is to reduce the complexity of a quantum circuit to accommodate many qubits.
The current and near-term quantum computers, known as noisy intermediate-scale
quantum (NISQ) devices, are characterized by low qubit counts, high gate error rates, and suffer from short qubit decoherence times~\cite{Preskill2018}.
Hence, optimizing quantum circuits into short-depth circuits is extremely important to successfully produce high-fidelity results on NISQ devices.

Quantum image processing (QIMP) extends the classical image processing operations to the quantum computing framework~\cite{yan2020quantum}. 
QIMP algorithms are used on images that have been represented in a quantum state. 
A variety of quantum image representation (QIR) methods has been developed~\cite{Yan2016}. 
The flexible representation of quantum images (FRQI)~\cite{Le2011,Le2011b},
the improved flexible representation of quantum images (IFRQI)~\cite{Khan2019}, 
the novel enhanced quantum representation (NEQR)~\cite{Zhang2013neqr},
the improved novel enhanced quantum representation (INEQR)~\cite{Jiang2015a},
the multi-channel representation of quantum images (MCRQI/MCQI)~\cite{Sun2011,Sun2013},
the novel quantum representation of color digital images (NCQI)~\cite{Sang2016}, and
the improved novel quantum representation of color digital images (INCQI)~\cite{Su2021}
are among the most powerful existing QIR methods.
These QIR methods became extremely popular due to two main factors.
First, their flexibility in encoding the positions and colors in a normalized quantum
state.
Second, image processing operations can be performed simultaneously on all pixels in the image by exploiting the superposition phenomenon of quantum mechanics.

In this paper, we introduce a uniform framework called the
\emph{quantum pixel representation} (QPIXL) that overarches all previously
mentioned quantum image representations and probably many more.
Furthermore, we propose a novel technique for preparing QPIXL representations
that requires fewer quantum gates for all the different representations,
compared to earlier results, and without introducing ancilla qubits.
The proposed method makes use of an efficient synthesis technique for the
uniformly controlled rotations~\cite{Mottonen2004} and uses only $\RY$ gates and
controlled-NOT ($\CX$) gates, making the resulting circuits practical in the NISQ era.
For example, the original FRQI state preparation method~\cite{Le2011} for an image with
$N = 2^n$ grayscale pixels uses $n+1$ qubits in total,
i.e., $n$ qubits for encoding the position and 1 qubit for the color,
and has a $\bigO(N^2)$ gate complexity.
Recently, the FRQI gate complexity has been reduced to $\bigO(N \log_2N)$ at the
price of introducing several extra ancilla qubits~\cite{Khan2019}.
In contrast, our QPIXL method for preparing an FRQI state has only a gate
complexity of $\bigO(N)$ and does not require extra ancilla qubits.
Additionally, we introduce a compression strategy to further reduce the gate
complexity of QPIXL representations.
In our experiments, the compression algorithm allows us to further reduce 
the gate complexity by up to 90\% without significantly sacrificing image
quality.
An implementation of our algorithms is publicly available as part of the
Quantum Image Pixel Library (\texttt{QPIXL++}) \cite{qpixlpp}
at \url{https://github.com/QuantumComputingLab}.
\texttt{QPIXL++} is build based on \texttt{QCLAB++} \cite{qclab,qclabpp}, which allows for creating and representing quantum circuits.

The remainder of the paper is organized as follows.
\reflitreview\ discusses the relevant related work in the literature. 
\refqpixl\ defines the quantum pixel representation formalism
that encompasses different image representations with specific color mappings.
\reffrqi\ reviews the FRQI format for grayscale image data as a specific
case of a quantum pixel representation, and its circuit design using
pixel-by-pixel multi-controlled rotation gates~\cite{Le2011}.
In \refourfrqi, we introduce the uniformly controlled rotation approach
as a replacement of the pixel-by-pixel one and show that it improves the gate
complexity to all earlier results~\cite{Le2011,Khan2019}.
\refcompr\ introduces a novel compression technique to further reduce the
gate complexity of the uniformly controlled rotation circuit.
In \refcolor, we extend our formalism to MCRQI for color images, and to IFRQI and NEQR formats, which have various advantages compared to FRQI.
We show how a compressed uniformly controlled rotation circuit can be used
to improve the gate complexity for all these representations.
\refexamples\ provides simulated examples obtained with \texttt{QPIXL++}
that demonstrate the significant reduction in gate complexity obtained with our
method and the efficacy of the compression algorithm.
Finally, \refconclude\ summarizes the proposed work and discusses
future research.


\section{\seclitreview}
\label{sec:litreview}

Almost every image processing algorithm~\cite{Gonzales2018} developed in the classical sense can also be developed in the quantum environment. 
These quantum versions may be computationally faster and may handle data more effectively by taking advantage of properties such as coherence, superposition, and entanglement associated with quantum science.
How an image is represented on a quantum computer dramatically
influences the image processing operations that can be applied.
Hence, QIR has become a vital area of study in QIMP.
Early approaches are the qubit lattice representation~\cite{Venegas-Andraca2003}
and the flexible representation of quantum images (FRQI)~\cite{Le2011}. The latter, which is the FRQI method forms the foundation of our work.
The former is a quantum counterpart of classical image representation models
without any significant performance improvement, while FRQI is based on
quantum mechanical phenomena and captures both the color and geometry of an
image in one quantum state.
Besides its flexibility and the use of fewer qubits, FRQI can also perform both geometric and color operations on the image concurrently~\cite{Su2020}.

Since the FRQI only uses one qubit for storing the color information,
the number of measurements to accurately retrieve an image can be very large.
The NEQR addresses this issue by storing the color
information in orthogonal states allowing for color retrieval in a single measurement.
Although the NEQR allows for accurate image retrieval, it requires significantly
more qubits and does not utilize the superposition principle in the color
qubit sequence, i.e., $\ell$ qubits basis states are used for images with bit depth $\ell$.
On the other hand, the IFRQI combines both ideas and utilizes limited and discrete 
levels of superposition that are maximally distinguishable.
The IFRQI therefore ensures accurate image retrieval with a small number of measurements,
however, it requires $\log_2(N) - 2$ extra ancilla qubits.
Other existing quantum image representation models are
the quantum image representation for log-polar images (QUALPI)~\cite{Zhang2013qualpi},
the $n$-qubit normal arbitrary superposition state (NASS)~\cite{LI2014212}, and
the generalized quantum image representation (GQIR)~\cite{Jiang2015b}.

Several quantum image processing algorithms have been introduced in the literature using these QIRs.
For example, Zhang et al.~\cite{Zhang2015,Zhang2015_2} introduced an image edge extraction algorithm (QSobel) based on FRQI and also a quantum feature extraction framework based on NEQR.
Jiang et al.~\cite{Jiang2019} recently proposed a new quantum image median filtering based on the NEQR.
There are image segmentation algorithms that utilizes different QIRs along with the quantum Fourier transform~\cite{Nielsen:2011:QCQI,Camps2020}.
Jiang et al.~\cite{Jiang2015b} developed a new quantum image scaling up algorithm based on the GQIR.
Li et al.~\cite{Li2018} developed a quantum version of the wavelet packet transforms based on the NASS.
Zhou et al.~\cite{Zhou2017} proposed a quantum realization of the bilinear interpolation method for NEQR.
There are several other examples in major application areas including image filtering~\cite{Caraiman2013,Yuan2018,Li2018_2,Yuan2017}, image segmentation~\cite{Caraiman2014,Caraiman2015,Li2020}, and machine learning~\cite{Nakaji2021,Huang2021,Abbas2021,Biamonte2017,Cong2019}.

In order to run a quantum algorithm on a NISQ device, it first needs to be
synthesized into elementary 1- and 2-qubit gates.
The original implementation of the FRQI~\cite{Le2011} required $\bigO(N^2)$
elementary gates, while the more recent implementation by Khan~\cite{Khan2019}
reduced the complexity to $\bigO(64 N \log_2 N)$ elementary gates
by introducing $\log_2(N) - 2$ extra ancilla qubits.
We propose a novel QPIXL synthesis approach that reduces the FRQI gate complexity
to $\bigO(2N)$, i.e., $N$ rotation $R_y$ gates and $N$ $\CX$ gates,
and does not require ancillary qubits.
Furthermore, our QPIXL synthesis approach also reduces the original IFRQI gate
complexity from $\bigO(p N \log_2 N)$ to only $\bigO(pN)$
and also gets rid of the ancilla qubits.
Similar gains are obtained for preparing (I)NEQR, MCRQI, and (I)NCQI states.


\section{\secqpixl}
\label{sec:quantum-image}

Some of the most widely used representations for quantum images, such as
(I)FRQI~\cite{Le2011,Khan2019},
(I)NEQR~\cite{Zhang2013neqr,Jiang2015a},
MCRQI~\cite{Sun2013}, and
(I)NCQI~\cite{Sang2016,Su2021},
can all be described by the following general definition for quantum image representations.
This representation is similar to the pixel representation for images on traditional computers and captures both pixel colors and positions into a single quantum state $\ket{I}$ that we call a quantum pixel representation, QPIXL in short.

\begin{definition}[Square QPIXL]
\label{def:QI}
The quantum state for the QPIXL representation of
a $2^m \times 2^m$ image $P = \left[p_{ij}\right]$, where
each pixel $p_{ij}$ has color $c_{ij}$,
is given by the normalized state%
\footnote{We remark that the order of $\ket{k}$ and $\ket{c_k}$ in \cref{def:QI}
is reversed compared to the original definition~\cite{Le2011,Le2011b}.
Our ordering is consistent with the quantum circuit implementation for $\ket{I}$
provided in~\refqpixlqc\ and in the original work~\cite{Le2011,Le2011b}.}
\begin{equation}
\ket{I} = \frac{1}{2^m} \sum_{k=0}^{2^{2m}-1} \ket{k} \otimes \ket{c_k},
\label{eq:QI}
\end{equation}
where $\ket{k}$ are the computational basis states on $2m$-qubits and
$\ket{c_k}$ is an encoding of the color information $c_{ij}$ in a quantum
state on one or more qubits.
The color values $\ket{c_k}$ should be regarded as a vectorized version of the 2D
color values $c_{ij}$, i.e., $\ket{c_k} \mapsto c_{ij}$ for $k = i + j \cdot 2^m$.
\end{definition}

We observe that the QPIXL state $\ket{I}$ creates an equal superposition over the computational basis states of the $2m$-qubits in the first register, which encodes the pixel positions, and applies a tensor product with the state on the second register that encodes the color information.
\Cref{def:QI} is general because it allows for flexibility in the type
of color information and color encoding that is used.
The mentioned QPIXL representations differ in their approach to map $c_{ij}$ to $\ket{c_k}$.

Since~\cref{def:QI} can trivially be extended to rectangular, 3D, and higher dimensional images, 
we will use the following more general definition.

\begin{definition}[General QPIXL]
\label{def:QI-gen}
The quantum state for the QPIXL representation of
an image of $N$ pixels $p_k$ 
is given by the normalized quantum state
\begin{equation}
\ket{I} = \frac{1}{\sqrt{2^n}} 
\left( \sum_{k=0}^{N-1} \ket{k} \otimes \ket{c_k} + \sum_{k=N}^{2^n-1} \ket{k} \otimes \ket{0} \right),
\label{eq:QI-gen}
\end{equation}
where $n = \lceil{\log_2{N}\rceil}$, $\ket{c_k}$ is an encoding of the color information
of pixel $p_k$, 
and $\ket{k}$ are the computational basis states on $n$-qubits.
\end{definition}

Remark that in case the number of pixels $N$ is not a power of $2$, \cref{def:QI-gen} 
appends zero-valued pixels for $k = N,N+1,\ldots,2^{\lceil{\log_2{N}}\rceil}-1$.
Consequently, the state~\eqref{eq:QI-gen} is fully determined by the $N$ pixel values $p_k$.
Without loss of generality, we will assume that $N = 2^n$ in the remainder of the paper.

\subsection{\secqpixlqc}
\label{sec:qpixl-qc}

The preparation of a QPIXL state on a quantum computer can be considered as a state preparation procedure, i.e., $\ket{I}$ is the result of a quantum circuit $\UQPIXL$ applied to the all-zero state $\ket0^{\otimes n+\ell}$,
where $n$ qubits are used to encode the pixel position and $\ell$ qubits are used for the color information.
\[\small
\ket{0}^{\otimes n+\ell}\left\{\rule{0em}{1.25em}\right.
\begin{myqcircuit*}{0.25}{0.75}
\lstick{} & {/} \qw & \multigate{1}{\UQPIXL} & \qw  \\
\lstick{} & {/} \qw & \ghost{\UQPIXL}        & \qw
\end{myqcircuit*}
\left\}\rule{0em}{1.25em}\right.\ket{I}
\qquad = \qquad~
\ket{0}^{\otimes n+\ell}\left\{\rule{0em}{1.25em}\right.
\begin{myqcircuit*}{0.25}{0.75}
\lstick{} & {/} \qw & \gate{H^{\otimes n}}  & \multigate{1}{U_{\ket{c}}} & \qw \\
\lstick{} & {/} \qw & \qw & \ghost{U_{\ket{c}}} & \qw \\
\end{myqcircuit*}
\left\}\rule{0em}{1.25em}\right.\ket{I}
\]
All QPIXL states are prepared in two steps: first creating an
equal superposition over the $n$ qubits that determine the pixel positions
and afterwards adding the color information to the state by means
of a unitary $U_{\ket{c}}$.
In matrix notation, this procedure yields
\begin{equation}
\ket{I} = \UQPIXL \ket0^{\otimes n+\ell} = U_{\ket{c}} (H^{\otimes n} \otimes I^{\otimes \ell} )  \ket0^{\otimes n+\ell},
\label{eq:ketI-prep}
\end{equation}
where $H^{\otimes n} \otimes I^{\otimes \ell}$ creates an equal superposition over the first $n$ qubits:
\begin{equation}
( H^{\otimes n} \otimes I^{\otimes \ell} ) \ket0^{\otimes n+\ell}
 = H^{\otimes n} \ket0^{\otimes n} \otimes \ket0^{\otimes \ell}
 = \frac{1}{\sqrt{N}} \sum_{k=0}^{N-1} \ket{k} \otimes \ket{0}^{\otimes \ell}.
 \label{eq:superposition}
\end{equation}


\section{\secfrqi}
\label{sec:frqi}

The FRQI~\cite{Le2011,Le2011b} fits \Cref{def:QI,def:QI-gen} of the QPIXL framework and is applicable to grayscale image data. 
An FRQI encoding uses only 1 qubit for the pixel intensity information $\ket{c_k}$ with a color mapping that is defined as follows.

\begin{definition}[FRQI mapping]
\label{def:FRQI}
For a grayscale image of $N$ pixels $p_{k}$ where
each pixel has a grayscale value $g_{k} \in \left[0, K\right]$, i.e., 
an integer value between $0$ and the maximum intensity $K$,
the QPIXL state with the FRQI mapping $\ket{I_{\text{FRQI}}}$ is defined by \Cref{def:QI-gen} with
the color mapping used in \eqref{eq:QI-gen} given by 
\begin{align}
\ket{c_k} & = \cos(\theta_k) \ket{0} + \sin(\theta_k)\ket{1}, &
\theta_k & = \frac{\pi/2}{K} \, g_k,
\label{eq:ketci}
\end{align}
with
$\ket{0} = \left[\begin{smallmatrix} 1 \\ 0 \end{smallmatrix}\right]$ and
$\ket{1} = \left[\begin{smallmatrix} 0 \\ 1 \end{smallmatrix}\right]$.
\end{definition}

Observe that the FRQI representation of an $N$-pixel grayscale image requires 
$n +1$ qubits in total: $n$ qubits for the pixel positions in $\ket{k}$ and 
$1$ qubit for encoding the corresponding pixel intensity information in $\ket{c_k}$.
By~\cref{eq:ketci}, we have that $\theta_k \in \left[0,\frac{\pi}{2}\right]$
and
\begin{equation}
\ket{c_k} = \begin{bmatrix} \cos(\theta_k) \\ \sin(\theta_k) \end{bmatrix}.
\end{equation}

\cref{def:FRQI} is flexible because the grayscale value of each pixel $p_k$ can be encoded by choosing the angles $\theta_k$ accordingly.
For example, consider an 8-bit grayscale image where each pixel $p_k$ 
has a grayscale value $g_k$ between $0$ and $255$, then the angles $\theta_k$ in \cref{eq:ketci} are given by
\begin{equation}
\theta_k = \frac{\pi/2}{255} \, g_k.
\end{equation}
On the other hand, repeated measurement of the quantum state $\ket{c_k}$
yields the probabilities $\alpha_k^2 = \cos^2(\theta_k)$ and 
$\beta_k^2 = \sin^2(\theta_k)$ for the basis states $\ket0$ and $\ket1$, respectively.
Hence, we can retrieve the grayscale values from these measurements by
\begin{equation}
g_k = \frac{255}{\pi/2} \arctan\left(\frac{\beta_k}{\alpha_k}\right).
\end{equation}

We assume until~\refcolor\ that all image data is in grayscale and that we
use the FRQI encoding from \Cref{def:FRQI}.

\subsection{QPIXL-FRQI quantum circuit implementation}
\label{sec:frqi-qc}

The circuit structure introduced in~\refqpixlqc\ can be used to
prepare the FRQI state on a quantum computer.  In this case we have
$\ell = 1$ and $U_{\ket{c}}$ that implements the mapping from \Cref{def:FRQI},
we will denote this unitary as $\UR$.
This specification yields
\begin{equation}
\ket{I_\text{FRQI}} = \underbrace{\UR (H^{\otimes n} \otimes I)}_{\UFRQI} \ket0^{\otimes n+1},
\label{eq:ketI-FRQI-prep}
\end{equation}
with, according to \cref{eq:superposition},
\begin{align}
(H^{\otimes n} \otimes I) \ket0^{\otimes n+1}
 &= \frac{1}{\sqrt{N}} \sum_{k=0}^{N-1} \ket{k} \otimes \ket{0}  
  = \frac{1}{\sqrt{N}}
    {\underbrace{\begin{bmatrix} 1 & 1 & \cdots & 1 \end{bmatrix}}_N}\T
    \otimes \begin{bmatrix} 1 & 0 \end{bmatrix}\T,\\
 &= \frac{1}{\sqrt{N}}
    {\underbrace{\begin{bmatrix} 1 & 0 & 1 & 0 & \cdots & 1 & 0 \end{bmatrix}}_{2N}}\T.
\end{align}
We define
\begin{align}
\UR
 &= R_y(2\theta_0) \oplus R_y(2\theta_1) \oplus \cdots \oplus
    R_y(2\theta_{N-1})
  = \begin{bmatrix}
      R_y(2\theta_0) \\
      & R_y(2\theta_1) \\
      && \ddots \\
      &&& R_y(2\theta_{N-1})
    \end{bmatrix},
\end{align}
with
\begin{equation}
R_y(2\theta_i) =
\begin{bmatrix}
  \cos(\theta_i) & -\sin(\theta_i) \\
  \sin(\theta_i) &  \phantom{-}\cos(\theta_i)
\end{bmatrix}.
\end{equation}
Since $\UR$ is by definition a block diagonal matrix with $N$ $2 \times 2$ blocks and
\begin{equation}
R(2\theta_i) \begin{bmatrix} 1 \\ 0 \end{bmatrix}
 = \begin{bmatrix} \cos(\theta_i) \\ \sin(\theta_i) \end{bmatrix},
\end{equation}
the prepared FRQI state \eqref{eq:ketI-FRQI-prep} becomes
\begin{align}
\ket{I_\text{FRQI}} & = \frac{1}{\sqrt{N}}
\begin{bmatrix}
\cos(\theta_0) & \sin(\theta_0) &
\cos(\theta_1) & \sin(\theta_1) & \cdots &
\cos(\theta_{N-1}) & \sin(\theta_{N-1})
\end{bmatrix}\T,\\
&= 
\frac{1}{\sqrt{N}} \sum_{k=0}^{N-1} \ket{k} \otimes \ket{c_k},
\end{align}
which is a vector of length $2N$ holding the cosine and sine values of the angles of all the pixels.
It can be directly verified that this definition of $\UR$ agrees with \Cref{def:FRQI}.

We can implement the $\UR$ circuit on a quantum computer
by using $N$ multi-controlled $\RY$ gates~\cite{Le2011}.
We use the notation $C^n(\RY)$ for an $\RY$ gate with $n$ control qubits.
To illustrate this, we consider the FRQI encoding of a $2 \times 2$ image.
This 4 pixels image can be implemented as follows using 3 qubits and 4 $C^2(\RY)$ gates:
\[\small{
\begin{myqcircuit}
\lstick{\ket{0}} & \gate{H} & \qw\barrier[-2.5em]{2} &
         \ctrlo1 & \qw & \ctrlo1 & \qw &  \ctrl1 & \qw & \ctrl{1} & \qw \\
\lstick{\ket{0}} & \gate{H} & \qw &
         \ctrlo1 & \qw &  \ctrl1 & \qw & \ctrlo1 & \qw & \ctrl{1} & \qw &
         \rstick{\!\!\!\left\}\rule{0em}{2.25em}\right.\ket{I_\text{FRQI}}} \\
\lstick{\ket{0}} & \qw & \qw &
         \gate{R_y(2\theta_0)} & \qw & \gate{R_y(2\theta_1)} & \qw &
         \gate{R_y(2\theta_2)} & \qw & \gate{R_y(2\theta_3)} & \qw
\end{myqcircuit}
}
\]
The angles $\theta_i$ correspond to the pixel values $p_i$ for $i = 0,1,2,3$
according to \cref{eq:ketci}.
The decomposition of the block diagonal matrix $\UR$ into multi-controlled $R_y$ gates
corresponds to the following matrix decomposition
\begin{equation*}
\begin{bmatrix}
      I \\
      & I \\
      && I \\
      &&& R_y(2\theta_3)
\end{bmatrix}
\begin{bmatrix}
      I \\
      & I \\
      && R_y(2\theta_2) \\
      &&& I
\end{bmatrix}
\begin{bmatrix}
      I \\
      & R_y(2\theta_1) \\
      && I \\
      &&& I
\end{bmatrix}
\begin{bmatrix}
      R_y(2\theta_0) \\
      & I \\
      && I \\
      &&& I
\end{bmatrix},
\end{equation*}
where each multi-controlled gate sets a single $2 \times 2$ block on the diagonal.

In order to actually run the $\UFRQI$ circuit on a quantum computer, we need to further synthesize the multi-controlled $R_y$ gates into elementary 1- and 2-qubit gates. 
For the case of $C^2(R_y)$ gates this can be done as follows~\cite{Barenco1995}:
\[\small
\begin{myqcircuit}
 & \ctrl1 & \qw &&           && \ctrl2 & \ctrl1 & \qw & \ctrl1 & \qw & \qw \\
 & \ctrl1 & \qw && \push{=~} && \qw & \targ & \ctrl1 & \targ & \ctrl1 & \qw \\
 & \gate{R_y(2\theta_i)} & \qw && && \gate{R_y(\theta_i)} & \qw & \gate{R_y(-\theta_i)} & \qw & \gate{R_y(\theta_i)} & \qw
\end{myqcircuit}
\]
yielding the following $\UFRQI$ circuit for the 4 pixels image example:
\begin{align*}
\small{
\begin{myqcircuit}
\lstick{\ket{0}} & \gate{H} & \qw\barrier[-1.25em]{2} & \gate{X} &
  \ctrl2 & \ctrl1 & \qw    & \ctrl1 & \qw    & \qw\barrier[-1.25em]{2} & \qw &
  \ctrl2 & \ctrl1 & \qw    & \ctrl1 & \qw    & \qw \\
\lstick{\ket{0}} & \gate{H} & \qw & \gate{X} &
  \qw    & \targ  & \ctrl1 & \targ  & \ctrl1 & \qw & \gate{X} &
  \qw    & \targ  & \ctrl1 & \targ  & \ctrl1 & \qw \\
\lstick{\ket{0}} & \qw & \qw & \qw &
  \gate{R_y( \theta_0)} & \qw &
  \gate{R_y(-\theta_0)} & \qw &
  \gate{R_y( \theta_0)} & \qw & \qw &
  \gate{R_y( \theta_1)} & \qw &
  \gate{R_y(-\theta_1)} & \qw &
  \gate{R_y( \theta_1)} & \qw
\end{myqcircuit}} \\[5pt]
\small{%
\begin{myqcircuit}
& \qw\barrier[-1.25em]{2} & \gate{X} &
  \ctrl2 & \ctrl1 & \qw    & \ctrl1 & \qw    & \qw\barrier[-1.25em]{2} & \qw &
  \ctrl2 & \ctrl1 & \qw    & \ctrl1 & \qw    & \qw \\
& \qw & \gate{X} &
  \qw    & \targ  & \ctrl1 & \targ  & \ctrl1 & \qw & \gate{X} &
  \qw    & \targ  & \ctrl1 & \targ  & \ctrl1 & \qw &
  \rstick{\!\!\!\left\}\rule{0em}{2.25em}\right.\ket{I_\text{FRQI}}} \\
& \qw & \qw &
  \gate{R_y( \theta_2)} & \qw &
  \gate{R_y(-\theta_2)} & \qw &
  \gate{R_y( \theta_2)} & \qw & \qw &
  \gate{R_y( \theta_3)} & \qw &
  \gate{R_y(-\theta_3)} & \qw & 
  \gate{R_y( \theta_3)} & \qw
\end{myqcircuit}}
\end{align*}
By further decomposing the $C^1(R_y)$ gates into $3$ $R_y$ and
$2$ CNOT gates as follows~\cite{Nielsen:2011:QCQI},
\begin{align*}
\small{
\begin{myqcircuit}
& \ctrl1 & \qw &&\\
& \gate{R_y(\theta_i)} & \qw &&
\end{myqcircuit}
=
\begin{myqcircuit}
&&& \qw & \ctrl1 & \qw & \ctrl1 & \qw  & \qw \\
&&& \gate{R_y(\theta_i)} & \targ & \gate{R_y(-\nicefrac{\theta_i}{2})} & \targ & \gate{R_y(-\nicefrac{\theta_i}{2})} & \qw 
\end{myqcircuit}
}
\end{align*}
the directly implementable quantum circuit for $\UFRQI$ requires
44 single-qubit and 32 CNOT gates in total.
In the general case for images with $N = 2^n$ pixels, 
every individual pixel value is encoded by a $C^n(R_y)$ gate. 
Decomposing these gates into 1- and 2-qubit
gates by the method of Barenco et al.~\cite{Barenco1995} requires $\bigO(N)$ gates for every $C^n(R_y)$ gate.
This results in an overall circuit complexity for $\UFRQI$ that scales quadratically in $N$, i.e., $\bigO(N^2)$ elementary gates are required to implement the full $\UFRQI$ circuit for an $N$ pixels image on a quantum computer~\cite{Le2011}.
Khan~\cite{Khan2019} recently improved the asymptotic 
complexity to $\bigO(N \log_2 N)$ by using $n - 2$ ancilla qubits.


\section{\secourfrqi}
\label{sec:ourfrqi}

The complexity of implementing $\UFRQI$ is determined by the complexity of the circuit for $\UR$, a block diagonal matrix with $2 \times 2$ blocks corresponding to the pixel values.
In this section, we derive an alternative circuit implementation for $\UR$ that requires quadratically fewer gates compared to the method proposed by Le et al.~\cite{Le2011}, i.e., the asymptotic complexity of our novel implementation requires only $\bigO(N)$ quantum gates for a $N$-pixel image.
Our new approach thus has optimal asymptotic scaling.
It is also logarithmically faster compared to the method proposed by Khan~\cite{Khan2019} and requires no ancilla qubits.

We start by reviewing a special case of the method introduced by
M\"ott\"onen et al.~\cite{Mottonen2004} to implement a block diagonal matrix in a quantum circuit.
In that work, these circuits are called \emph{uniformly controlled $\RY$ rotations} because they uniformly use all
possible computational basis states in the control register.
Let us define the nomenclature and diagrammatic notation for uniformly controlled
$\RY$ rotations.
\begin{definition}[Uniformly controlled $\RY$ rotations]
Given $\bftheta \inR^N$, a vector of rotation angles, the uniformly controlled $\RY$ rotation
is defined as
\begin{equation}
    \UR
 = R_y(\theta_0) \oplus R_y(\theta_1) \oplus \cdots \oplus
    R_y(\theta_{N-1}),
\end{equation}
and represented diagrammatically as
\[\small
\begin{myqcircuit*}{0.25}{0.75}
\lstick{} & {/} \qw & \multigate{1}{\UR} &  \qw \\
\lstick{} & \qw &        \ghost{\UR} &  \qw
\end{myqcircuit*}
\quad = \quad~
\begin{myqcircuit}
& {/} \qw & \ctrlsq{1} & \qw &&\\
& \qw & \gate{R_y(\bftheta)} & \qw &&
\end{myqcircuit}
\]
The dashed line indicates the $n \ = \log_2(N)$ qubits
required for controlling the different diagonal positions in $\UR$.
The diagram on the right hand side uses a square control node to indicate that it
is uniformly controlled by the first $n$ qubits.
\end{definition}

We know from the previous section that we can implement $\UR$ by using $N$ $C^n(\RY)$
gates. Here, we show that we can do this more efficiently by using a circuit that only
consists of $\RY$ and $\CX$ gates.
As an illustrative example, let us consider the following circuit for 4 arbitrary angles $\hat \theta_0, \ldots, \hat \theta_3$:
\begin{align*}
\small
{\begin{myqcircuit}
	& \qw                       & \qw & \qw      & \qw & \qw                       & \qw & \ctrl{2} & \qw & \qw                       & \qw & \qw      & \qw & \qw                       & \qw & \ctrl{2} & \qw \\
	& \qw                       & \qw & \ctrl{1} & \qw & \qw                       & \qw & \qw      & \qw & \qw                       & \qw & \ctrl{1} & \qw & \qw                       & \qw & \qw      & \qw \\
	& \gate{R_y(\hat \theta_0)} & \qw & \targ    & \qw & \gate{R_y(\hat \theta_1)} & \qw & \targ    & \qw & \gate{R_y(\hat \theta_2)} & \qw & \targ    & \qw & \gate{R_y(\hat \theta_3)} & \qw & \targ    & \qw
\end{myqcircuit}}
\end{align*}
The following two properties of $R_y$ rotations are immediate:
\begin{align*}
R_y(\theta_0) \, R_y(\theta_1) & = R_y(\theta_0 + \theta_1), \\
X \, R_y(\theta) \, X & = R_y(-\theta),
\end{align*}
where $X$ is a NOT gate $\left[\begin{smallmatrix} 0 & 1 \\ 1 & 0 \end{smallmatrix} \right]$ that appears in the CNOT gates above.
We can analyze the circuit above using these two simple properties and show that the circuit does create a block diagonal matrix with $2 \times 2$ blocks on the diagonal: the $R_y$ rotations on the 3rd qubit are all block diagonal matrices and the CNOT gates permute some of the blocks depending on the index of the first two control qubits. 
If we list the four $2 \times 2$ diagonal blocks in binary order, or equivalently the state of the 1st and 2nd qubit, we see that 
the circuit has the following effect on each block:
\begin{align*}
00&: & \phantom{X} R_y(\hat \theta_3) \phantom{X} R_y(\hat \theta_2) \phantom{X}  R_y(\hat \theta_1) \phantom{X}  R_y(\hat \theta_0) & = R_y( \phantom{-}\hat\theta_3 + \hat\theta_2 + \hat\theta_1 + \hat\theta_0),\\
01&: & \phantom{X} R_y(\hat \theta_3) X R_y(\hat \theta_2) \phantom{X}  R_y(\hat \theta_1) X R_y(\hat \theta_0) & = R_y( \phantom{-}\hat\theta_3 - \hat\theta_2 - \hat\theta_1 + \hat\theta_0),\\
10&: & X R_y(\hat \theta_3) \phantom{X} R_y(\hat \theta_2) X R_y(\hat \theta_1) \phantom{X}  R_y(\hat \theta_0) & = R_y( - \hat\theta_3 - \hat\theta_2 + \hat\theta_1 + \hat\theta_0),\\
11&: & X R_y(\hat \theta_3) X R_y(\hat \theta_2) X R_y(\hat \theta_1) X R_y(\hat \theta_0) & = R_y( - \hat\theta_3 + \hat\theta_2 - \hat\theta_1 + \hat\theta_0).
\end{align*}
To implement a block diagonal matrix with this circuit, where the angles of the $R_y$ blocks correspond to $(\theta_0, \ldots, \theta_3)$, we get that the angles have to satisfy
\begin{align*}
\begin{bmatrix}
\theta_0 \\ \theta_1 \\ \theta_2 \\ \theta_3
\end{bmatrix}
=
\begin{bmatrix}
1 & \phantom{-}1 & \phantom{-}1 & \phantom{-}1 \\
1 & -1           & -1           & \phantom{-}1 \\
1 & \phantom{-}1 & -1           & -1           \\
1 & -1           & \phantom{-}1 & -1
\end{bmatrix}
\begin{bmatrix}
\hat\theta_0 \\ \hat\theta_1 \\ \hat\theta_2 \\ \hat\theta_3
\end{bmatrix}.
\end{align*}
This is a linear system with a specific structure, that we can rewrite as
\begin{equation}
\begin{bmatrix}
\theta_0 \\ \theta_1 \\ \theta_2 \\ \theta_3
\end{bmatrix}
=
\begin{bmatrix}
1 & \phantom{-}1 & \phantom{-}1 & \phantom{-}1 \\
1 & -1           & \phantom{-}1           & -1 \\
1 & \phantom{-}1 & -1           & -1           \\
1 & -1           & -1 & \phantom{-}1
\end{bmatrix}
\begin{bmatrix}
1 &   &   &   \\
  & 1 &   &   \\
  &   & 0 & 1 \\
  &   & 1 & 0
\end{bmatrix}
\begin{bmatrix}
\hat\theta_0 \\ \hat\theta_1 \\ \hat\theta_2 \\ \hat\theta_3
\end{bmatrix}
= 
( \hat H \otimes \hat H ) P_G
\begin{bmatrix}
\hat\theta_0 \\ \hat\theta_1 \\ \hat\theta_2 \\ \hat\theta_3
\end{bmatrix},
\label{eq:sys}
\end{equation}
where $\hat H = \left[\begin{smallmatrix} 1 & \phantom{-}1 \\ 1 & -1 \end{smallmatrix}\right]$ is a scaled version of the Hadamard gate and $P_G$ is the permutation matrix that transforms binary ordering to Gray code ordering.

It follows that, if we solve the linear system~\eqref{eq:sys} for 
$(\hat \theta_0, \ldots, \hat \theta_3)$, we can implement $\UR$
for any $2 \times 2$ image with only $8$ elementary gates: $4$ $R_y$
rotations and $4$ CNOT gates.
The $\UR$ circuit for the $2 \times 2$ example in the previous section
required $74$ gates: $42$ 1-qubit and $32$ CNOT gates.
Indeed, we have a quadratic improvement in gate complexity.

This strategy generalizes to block diagonal matrices $\UR$ 
that have $2^n$ $R_y$ blocks on their diagonal~\cite{Mottonen2004}.
The circuit structure consists of a sequence of length $2^n$ alternating between $R_y$ gates and CNOT gates. The $R_y$ gates act on the $(n+1)$st qubit, and thus correspond to block diagonal matrices with $2 \times 2$ blocks. The target qubit of the CNOT gates is set to the $(n+1)$st qubit and the control qubit for the $\ell$th CNOT gate is set to the bit where the $\ell$th and $(\ell + 1)$st Gray code differ.
If $\UR$ is determined by the angles $\bftheta = (\theta_0, \ldots, \theta_{2^n-1})$,
the angles of the circuit $\hat \bftheta = (\hat \theta_0, \ldots, \hat \theta_{2^n-1})$
can be computed through the linear system:
\begin{align}
\left( \hat H^{\otimes n} \, P_G \right) \hat \bftheta = \bftheta.
\label{eq:ls}
\end{align}
In \texttt{QPIXL++}, Eq.~\eqref{eq:ls} is solved with a matrix-free approach: the Gray permutation $P_G$ is performed in place and the scaled Walsh-Hadamard transform $\hat H^{\otimes n}$ is
implemented through a variant of the fast Walsh-Hadamard transform which requires $\bigO( N \log N)$ operations. 
This approach scales to large-scale images and we report some results in~\refexamples.

Our new $\UR$ circuit requires only $N$ $R_y$ rotation and $N$ CNOT gates for an image with $N$ pixels.
As this scales linearly in the number of pixels, the asymptotic complexity of our approach is optimal.
This is a quadratic improvement compared to the approach proposed by Le et al.~\cite{Le2011} that we described in~\reffrqi. 
The asymptotic complexities of both approaches are summarized in~\Cref{tab:complexities}.
We remark that as we require just 2 gates for every pixel, our constant prefactor is also considerably smaller compared to the works by Le et al.~\cite{Le2011} and Khan~\cite{Khan2019}.

\begin{table}[hbtp]
\centering\small%
\begin{tabularx}{0.75\columnwidth}{l|CCC}
\toprule
FRQI & Gate complexity & Ancilla qubits & Total qubits \\
\midrule
Le et al.~\cite{Le2011} & $\bigO(N^2)$        &   0   & $n+1$ \\
Khan~\cite{Khan2019}    & $\bigO(N \log_2 N)$ & $n-2$ & $2n-1$ \\
QPIXL                   & $\bigO(N)$          &   0   & $n+1$ \\
\bottomrule
\end{tabularx}
\caption{Summary of gate complexities and qubit count for preparing the FRQI state $\ket{I_\text{FRQI}}$
for an image with $N = 2^n$ pixels with the approaches of Le et al.~\cite{Le2011} and Khan~\cite{Khan2019}
compared to our method.}
\label{tab:complexities}
\end{table}


\section{\seccompr}
\label{sec:compr}

The proposed implementation of $\UR$ as presented in~\refourfrqi\ lends itself to an efficient circuit and thus image compression technique.
As an example, we describe this idea for an FRQI image with 8 pixels.

Assume that the FRQI angle representation of an image is given by
the vector $\bftheta \inR^8$ and that we have computed the transformed
vector $\hat \bftheta \inR^8$ according to~\cref{eq:ls}.
The coefficients of $\hat\bftheta$ are then used in the following circuit for $\UR$:
\begin{align*}
\small
{\begin{myqcircuit}
& \qw                       & \qw      & \qw                  & \qw      & \qw                  & \qw      & \qw                  &  \ctrl{3} & \qw                       & \qw      & \qw                  & \qw      & \qw                  & \qw      & \qw                  &  \ctrl{3}      & \qw\\
& \qw                       & \qw      & \qw                  & \ctrl{2} & \qw                  & \qw      & \qw                  &  \qw      & \qw                       & \qw      & \qw                  & \ctrl{2} & \qw                  & \qw      & \qw                  &  \qw & \qw \\
& \qw                       & \ctrl{1} & \qw                  & \qw      & \qw                  & \ctrl{1} & \qw                  &  \qw      & \qw                       & \ctrl{1} & \qw                  & \qw      & \qw                  & \ctrl{1} & \qw                  &  \qw      & \qw \\
& \gate{\hat \theta_0}      & \targ    & \gate{\hat \theta_1} & \targ    & \gate{\hat \theta_2} & \targ    & \gate{\hat \theta_3} &  \targ    & \gate{\hat \theta_4}      & \targ    & \gate{\hat \theta_5} & \targ    & \gate{\hat \theta_6} & \targ    & \gate{\hat \theta_7} &  \targ    & \qw
\end{myqcircuit}}
\end{align*}

For conciseness, we omit the $R_y$ labels and only state the rotation angle for the $R_y$ gates.
Now assume that the image after the permuted Walsh-Hadamard transform is of the form $\hat \bftheta = ( \hat\theta_0, \hat\theta_1, \delta, \delta, \delta, \delta, \delta, \hat\theta_7)$,
where $\delta$ are angles that can be considered negligible according to some compression criterion. 
A good approximation of the image is then given by
$\hat \bftheta = ( \hat\theta_0, \hat\theta_1, 0, 0, 0, 0, 0, \hat\theta_7)$.
This corresponds to the circuit below on the left where all $R_y$ rotations that have $0$ angle after compression have been removed.
This step results in a sequence of consecutive $\CX$ gates all with the same target qubit and different control qubits. 
All these $\CX$ gates commute with each other, so we can place
them in arbitrary order. 
Furthermore, two consecutive $\CX$ gates that have the same control qubit cancel
each other since their product is the identity. 
The circuit below on the left has in the middle 1 CNOT with the first
qubit as control, 2 $\CX$s with the second qubit as control that cancel out, and 3 $\CX$s with
the third qubit as control of which two cancel with each other. 
It follows that the circuit on the left is equivalent to the circuit on the right with the
redundant $\CX$ gates removed.
\begin{align*}
\small
{\begin{myqcircuit}
& \qw                       & \qw      & \qw                  & \qw      & \qw      &  \ctrl{3} & \qw      & \qw      & \qw      & \qw                  &  \ctrl{3}      & \qw\\
& \qw                       & \qw      & \qw                  & \ctrl{2} & \qw      &  \qw      & \qw      & \ctrl{2} & \qw      & \qw                  &  \qw & \qw \\
& \qw                       & \ctrl{1} & \qw                  & \qw      & \ctrl{1} &  \qw      & \ctrl{1} & \qw      & \ctrl{1} & \qw                  &  \qw      & \qw \\
& \gate{\hat \theta_0}      & \targ    & \gate{\hat \theta_1} & \targ    & \targ    &  \targ    & \targ    & \targ    & \targ    & \gate{\hat \theta_7} &  \targ    & \qw
\end{myqcircuit}}
\qquad\Longrightarrow\qquad
{\begin{myqcircuit}
& \qw                       & \qw      & \qw                  & \ctrl{3} &  \qw      & \qw                  &  \ctrl{3}      & \qw\\
& \qw                       & \qw      & \qw                  & \qw      &  \qw      & \qw                  &  \qw & \qw \\
& \qw                       & \ctrl{1} & \qw                  & \qw      &  \ctrl{1} & \qw                  &  \qw      & \qw \\
& \gate{\hat \theta_0}      & \targ    & \gate{\hat \theta_1} & \targ    &  \targ    & \gate{\hat \theta_7} &  \targ    & \qw
\end{myqcircuit}}
\end{align*}

As we describe next, this procedure easily generalizes to images of arbitrary size.
After having computed $\hat\bftheta$, apply a compression criterion to
set the negligible coefficients $\hat\theta_i$ to 0. 
Next, remove the corresponding $R_y$ rotations 
with 0 angle from the $\UR$ circuit. 
Finally, perform a parity check on the control qubits of
consecutive $\CX$s in the $\UR$ circuit: no $\CX$ is required for control qubits with even parity, one $\CX$ is required
for control qubits with odd parity.

This algorithm is implemented in~\texttt{QPIXL++}~\cite{qpixlpp}.
The compression criterion that we adopted selects
a fixed percentage of the coefficients $\hat\theta_i$ with largest magnitude and thus of most importance.
For example, a compression setting of $0\%$ retains all nonzero coefficients in $\hat\bftheta$, while a
compression of $40\%$ sets the $40\%$ smallest coefficients $|\hat\theta_i|$ to zero. 
As we show in~\refexamples,
this method can achieve high compression ratios while maintaining many features of the uncompressed image.
The advantage of our approach is that we can discard coefficients after the Walsh-Hadamard transformation
has been applied.
In this way nonlocal correlations can be approximated with fewer coefficients compared to the untransformed data which can allow for improved compressibility.


\section{\seccolor}
\label{sec:color}

In this section, we extend our novel circuit implementation for
$\UFRQI$ for grayscale data to different image representations that
fit in \Cref{def:QI,def:QI-gen}.
The key difference between all representations is the definition of the
color encoding in the quantum state $\ket{c_k}$
from \cref{def:QI-gen}.
As long as we express this color mapping in terms of a combination of $\RY$
rotations, we can use our compressed implementation for the uniformly
controlled $\RY$ rotations.

\subsection{IFRQI}

The improved FRQI method introduced by Khan~\cite{Khan2019} combines ideas from
the FRQI and NEQR representations. It improves upon the measurement problem for FRQI
by allowing for only 4 discrete superpositions that are maximally distinguishable upon projective
measurement in the computational basis.
The IFRQI color mapping for a grayscale image with bit depth $2p$ is defined as follows.

\begin{definition}[IFRQI mapping]
\label{def:IFRQI}
For a grayscale image of $N$ pixels where each pixel $p_k$
has a grayscale value $g_k \in [0, 2^{2p}-1]$ with binary
representation $b^0_k b^1_k \cdots b^{2p-1}_k$,
the IFRQI state $\ket{I_\text{IFRQI}}$ is defined by \Cref{def:QI-gen} 
with the color mapping used in \eqref{eq:QI-gen} given by
\begin{equation}
\ket{c_k} = \ket{c_k^0 c_k^1 \cdots c_k^{p-1}},
\label{eq:IFRQI-gs}
\end{equation}
where, for $i = 0, \cdots, p-1$
\begin{align*}
\ket{c_k^i} & = 
 \cos(\theta_k^i) \ket{0} + \sin(\theta_k^i)\ket{1}, &
 \theta_k^i & =
    \begin{cases}
    0, & \text{if } b^{2i}_k b^{2i+1}_k = 00 \\
    \frac{\pi}{5}, & \text{if } b^{2i}_k b^{2i+1}_k = 01 \\
    \frac{\pi}{2} - \frac{\pi}{5}, & \text{if } b^{2i}_k b^{2i+1}_k = 10 \\
    \frac{\pi}{2}, & \text{if } b^{2i}_k b^{2i+1}_k = 11
    \end{cases}.
\end{align*}
\end{definition}

We observe that the IFRQI mapping combines two bits of color information into
one rotation. It follows that for an image with bit-depth $2p$, we can prepare
$\ket{I_\text{IFRQI}}$ using the circuit presented in~\Cref{fig:circuits}(a) with $p$ uniformly controlled $\RY$ rotations.
The rotation angles $\bftheta^i$ correspond to bits $2i$ and $2i+1$ of all $N$ pixels
according to the values defined in \Cref{def:IFRQI}. These uniformly controlled rotations
can be compressed independently with our compression algorithm.
The gate and qubit complexites for IFRQI with our method compared to Khan~\cite{Khan2019}
are listed in \Cref{tab:complexities-all}.

\subsection{NEQR}

The idea for NEQR is to use a color mapping that directly encodes
the length $\ell$ bitstring for the grayscale information in the
computational basis states on $\ell$ qubits.
The NEQR states for different colors are thus orthogonal and can be distinguished
with a single projective measurement in the computational basis.
In our QPIXL framework, the NEQR mapping can be defined as follows.

\begin{definition}[NEQR mapping]
\label{def:NEQR}
For a grayscale image of $N$ pixels where each pixel $p_k$
has a value $g_k \in [0, 2^{\ell}-1]$ with binary
representation $b^0_k b^1_k \cdots b^{\ell-1}_k$,
the NEQR state $\ket{I_\text{NEQR}}$ is defined by \Cref{def:QI-gen} 
with the color mapping used in \eqref{eq:QI-gen} given by
\begin{equation}
\ket{c_k} = \ket{c_k^0 c_k^1 \cdots c_k^{\ell-1}},
\label{eq:NEQR-gs}
\end{equation}
where 
\begin{align*}
\ket{c_k^i} & = 
 \cos(\theta_k^i) \ket{0} + \sin(\theta_k^i)\ket{1}, &
 \theta_k^i & =
    \begin{cases}
    0, & \text{if } b^{i}_k = 0 \\
    \frac{\pi}{2}, & \text{if } b^{i}_k = 1
    \end{cases}.
\end{align*}
\end{definition}

By choosing the rotation angles $\theta^i_k$ orthogonal, we ensure
that the color information in $\ket{I_\text{NEQR}}$ can be retrieved through a single
projective measurement.
The NEQR state can be prepared through the circuit shown in~\Cref{fig:circuits}(b), where the uniformly
controlled rotations can again be compressed with our method.
The gate complexities for the uncompressed circuits are listed in \Cref{tab:complexities-all}.

\subsection{MCRQI}

If we want to extend the applicability of the FRQI from grayscale to color image data,
we have to allow for different color channels.
This approach was dubbed multi-channel representation of quantum images (MCRQI)~\cite{Sun2013}.
We adapt their definition for RGB image data to our formalism and make some minor
modifications.

\begin{definition}[MCRQI mapping]
\label{def:MCRQI}
For a color image of $N$ RGB pixels,
where the color of each pixel $p_k$ is given by an RGB triplet $(r_k,g_k,b_k) \in \left[0, K\right]$,
the MCRQI state $\ket{I_{\text{MCRQI}}}$ is defined by \Cref{def:QI-gen} with
the color mapping used in \eqref{eq:QI-gen} given by 
\begin{equation}
\ket{c_k} = \ket{r_k g_k b_k},
\label{eq:MCRQI}
\end{equation}
where 
\begin{align*}
\ket{r_k} & = \cos(\theta_k) \ket{0} + \sin(\theta_k)\ket{1}, &
\theta_k & = \frac{\pi/2}{K} \, r_k, \\
\ket{g_k} & = \cos(\phi_k) \ket{0} + \sin(\phi_k)\ket{1}, &
\phi_k & = \frac{\pi/2}{K} \, g_k, \\
\ket{b_k} & = \cos(\gamma_k) \ket{0} + \sin(\gamma_k)\ket{1}, &
\gamma_k & = \frac{\pi/2}{K} \, b_k.
\end{align*}
\end{definition}

We see that to encode the color information for an RGB image, we only require 2 additional
qubits compared to grayscale data, which is a significant improvement over the classical case.
Furthermore, we encode the color mapping as a tensor product of three qubit states, while Sun et al.~\cite{Sun2013} encodes the information in the coefficients of the color qubits,
which entangles their state.
Our implementation has the advantage that the different color channels are easily
treated separately, while the color information can still be retrieved thanks to the normalization constraint.

The circuit implementation of $\ket{I_{\text{MCRQI}}}$ for the RGB mapping defined in \Cref{def:MCRQI}
then simply combines three uniformly controlled rotation circuits with different target qubits and coefficient vectors determined by the respective color intensities as shown in~\Cref{fig:circuits}(c).
As the RGB color channels are independent of each other and the uniformly controlled
$\RY$ gates have different target qubits, each of them can be compressed separately.
The asymptotic gate complexity of our method compared to the work by Sun et al.~\cite{Sun2013} is listed in~\Cref{tab:complexities-all}. 
As that work essentially uses the construction of Le et al.~\cite{Le2011}, we obtain a quadratic improvement before compression.

\subsection{INCQI}

Similarly to the NEQR, the (I)NCQI uses a color mapping directly encoding the length $\ell$ bitstring for each color value in a RGB$\alpha$ image in the computational basis stated on $\ell$ qbits.
Consequently, this QIR can also be easily represented by our QPIXL framework through the mapping defined as follows.

\begin{definition}[INCQI mapping]
\label{def:NCQI}
For a color image of N RGB$\alpha$ pixels, where the color of each pixel $p_k$ is given by a tuple $(r_k,g_k,b_k,\alpha_k)$ and each channel value in the range $[0, 2^{\ell}-1]$ has a binary representation, the INCQI state $\ket{I_{\text{INCQI}}}$ is defined by \Cref{def:QI-gen} with the color mapping used in \eqref{eq:QI-gen} given by
\begin{equation}
\ket{c_k} = \ket{r_kg_kb_k\alpha_k} = \ket{r_k^0r_k^1\dots r_k^{\ell-1}g_k^0g_k^1\dots g_k^{\ell-1}b_k^0b_k^1\dots b_k^{\ell-1}\alpha_k^0\alpha_k^1\dots \alpha_k^{\ell-1}}
\label{eq:INCQI}
\end{equation}
where
\begin{align*}
\ket{r_k^i} & = 
 \cos(\theta_k^i) \ket{0} + \sin(\theta_k^i)\ket{1}, &
 \theta_k^i & =
    \begin{cases}
    0, & \text{if } b^{i}_k = 0 \\
    \frac{\pi}{2}, & \text{if } b^{i}_k = 1
    \end{cases}.\\
\ket{g_k^i} & = 
 \cos(\phi_k^i) \ket{0} + \sin(\phi_k^i)\ket{1}, &
 \phi_k^i & =
    \begin{cases}
    0, & \text{if } b^{i}_k = 0 \\
    \frac{\pi}{2}, & \text{if } b^{i}_k = 1
    \end{cases}.\\
\ket{b_k^i} & = 
 \cos(\gamma_k^i) \ket{0} + \sin(\gamma_k^i)\ket{1}, &
 \gamma_k^i & =
    \begin{cases}
    0, & \text{if } b^{i}_k = 0 \\
    \frac{\pi}{2}, & \text{if } b^{i}_k = 1
    \end{cases}.\\
\ket{\alpha_k^i} & = 
 \cos(\psi^i) \ket{0} + \sin(\psi_k^i)\ket{1}, &
 \psi_k^i & =
    \begin{cases}
    0, & \text{if } b^{i}_k = 0 \\
    \frac{\pi}{2}, & \text{if } b^{i}_k = 1
    \end{cases}.
\end{align*}
\end{definition}

\begin{figure}[hbtp]
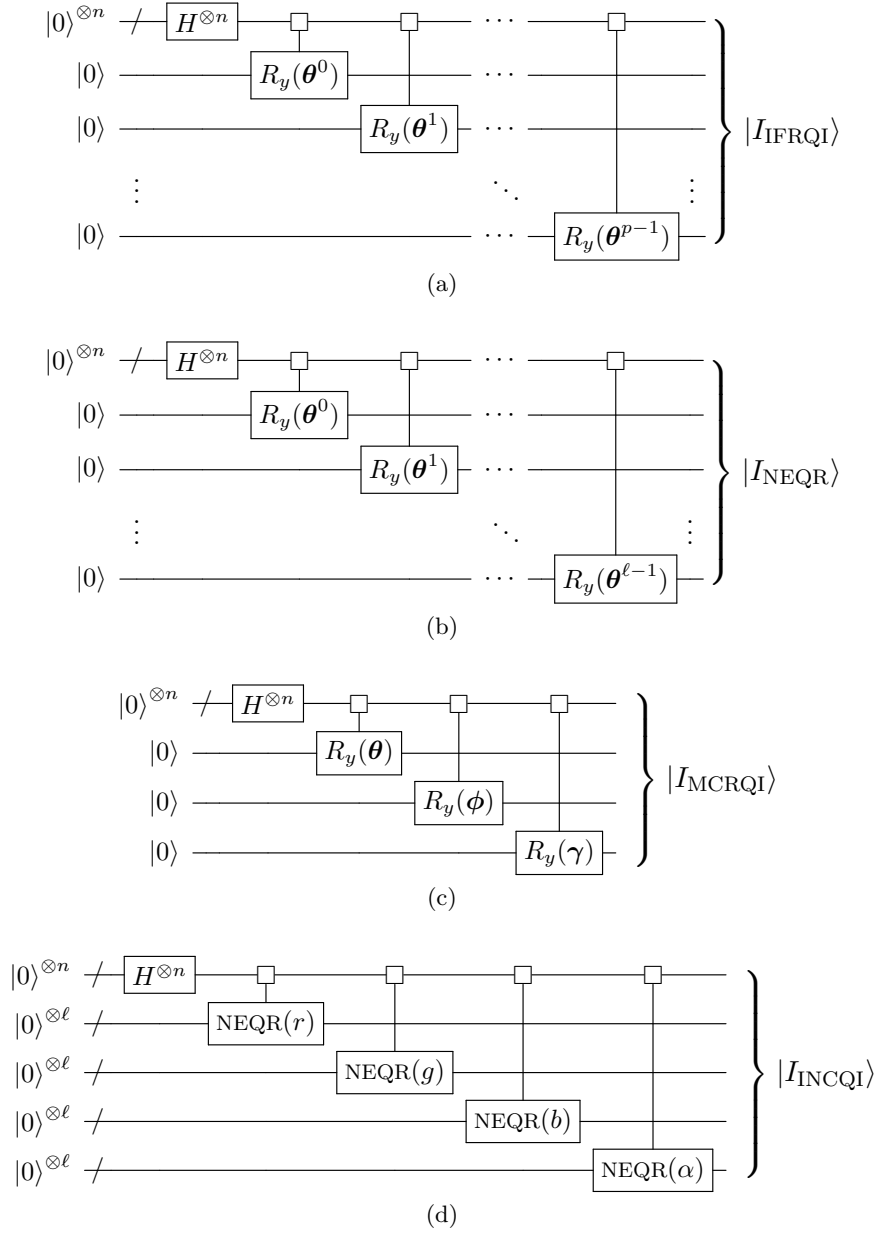

\centering
\subfloat[]{\qquad~~$\small
\begin{myqcircuit}
\lstick{\ket{0}^{\otimes n}} & {/} \qw & \qw & \gate{H^{\otimes n}} & \ctrlsq{1} & \ctrlsq{2} & \qw & \push{\cdots~} & \qw & \ctrlsq{4} &\qw & \qw \\
\lstick{\ket{0}} & \qw & \qw & \qw & \gate{R_y(\bftheta^0)} & \qw & \qw & \push{\cdots~} & \qw & \qw & \qw & \qw \\
\lstick{\ket{0}} & \qw & \qw & \qw & \qw & \gate{R_y(\bftheta^1)} & \qw & \push{\cdots~} & \qw & \qw & \qw & \qw \\
& \push{\vdots} & & & & & &  \push{\ddots} & & & \vdots \\
\lstick{\ket{0}} & \qw & \qw & \qw & \qw & \qw & \qw & \push{\cdots~} & \qw & \gate{R_y(\bftheta^{p-1})} & \qw & \qw  
\end{myqcircuit}
\left\}\rule{0em}{4.75em}\right.\ket{I_\text{IFRQI}}
$}\\[15pt]
\subfloat[]{\qquad~~$\small
\begin{myqcircuit}
\lstick{\ket{0}^{\otimes n}} & {/} \qw & \qw & \gate{H^{\otimes n}} & \ctrlsq{1} & \ctrlsq{2} & \qw & \push{\cdots~} & \qw & \ctrlsq{4} &\qw & \qw \\
\lstick{\ket{0}} & \qw & \qw & \qw & \gate{R_y(\bftheta^0)} & \qw & \qw & \push{\cdots~} & \qw & \qw & \qw & \qw \\
\lstick{\ket{0}} & \qw & \qw & \qw & \qw & \gate{R_y(\bftheta^1)} & \qw & \push{\cdots~} & \qw & \qw & \qw & \qw \\
& \push{\vdots} & & & & & &  \push{\ddots} & & & \vdots \\
\lstick{\ket{0}} & \qw & \qw & \qw & \qw & \qw & \qw & \push{\cdots~} & \qw & \gate{R_y(\bftheta^{\ell-1})} & \qw & \qw  
\end{myqcircuit}
\left\}\rule{0em}{4.75em}\right.\ket{I_\text{NEQR}}
$}\\[15pt]
\subfloat[]{\qquad~~~$\small
\begin{myqcircuit}
\lstick{\ket{0}^{\otimes n}} & {/} \qw & \qw & \gate{H^{\otimes n}} & \ctrlsq{1} & \ctrlsq{2} & \ctrlsq{3} &\qw &&\\
\lstick{\ket{0}} & \qw & \qw & \qw & \gate{R_y(\bftheta)} & \qw & \qw & \qw && \\
\lstick{\ket{0}} & \qw & \qw & \qw & \qw & \gate{R_y(\bfphi)} & \qw & \qw && \\
\lstick{\ket{0}} & \qw & \qw & \qw & \qw & \qw & \gate{R_y(\bfgamma)} & \qw &&
\end{myqcircuit}
\!\!\!\left\}\rule{0em}{3.75em}\right.\ket{I_\text{MCRQI}}
$}\\[15pt]
\subfloat[]{\qquad~~$\small
\begin{myqcircuit}
\lstick{\ket{0}^{\otimes n}} & {/} \qw & \qw & \gate{H^{\otimes n}} & \ctrlsq{1} & \ctrlsq{2} & \ctrlsq{3} & \ctrlsq{4} & \qw &&\\
\lstick{\ket{0}^{\otimes \ell}} & {/}\qw & \qw & \qw & \gate{\text{\scriptsize{NEQR}}(r)} & \qw & \qw & \qw & \qw && \\
\lstick{\ket{0}^{\otimes \ell}} & {/}\qw & \qw & \qw & \qw & \gate{\text{\scriptsize{NEQR}}(g)} & \qw & \qw & \qw && \\
\lstick{\ket{0}^{\otimes \ell}} & {/}\qw & \qw & \qw & \qw & \qw & \gate{\text{\scriptsize{NEQR}}(b)} & \qw & \qw && \\
\lstick{\ket{0}^{\otimes \ell}} & {/}\qw & \qw & \qw & \qw & \qw & \qw & \gate{\text{\scriptsize{NEQR}}(\alpha)} & \qw &&
\end{myqcircuit}
\!\!\!\left\}\rule{0em}{4.25em}\right.\ket{I_\text{INCQI}}
$}\\
\caption{Circuits for the preparation of the IFRQI, NEQR, MCRQI, and INCQI states, where the uniformly controlled rotations can be compressed with our method.}
\label{fig:circuits}
\end{figure}

The definition above applies very similarly to the NCQI~\cite{Sang2016}, only removing channel $\alpha$ from the equation. 
The INCQI state can be prepared through the circuit shown in~\Cref{fig:circuits}(d).
This circuit is built using an NEQR circuit for each channel of the ICNQI.
Similarly to previous QIRs, the uniformly controlled rotations used here can also be compressed with our method.
The gate complexities for the uncompressed circuits are listed in~\Cref{tab:complexities-all}.

\begin{table}[hbtp]
\centering\small%
\begin{tabularx}{\textwidth}{l|lcCC|cC}
\toprule
 & \multicolumn{4}{c|}{Literature} & \multicolumn{2}{c}{QPIXL} \\
Method & Reference & Gate complexity & Ancilla qubits & Total qubits & Gate complexity & Total qubits \\
\midrule
\multirow{2}{*}{FRQI} & Le et al.~\cite{Le2011} & $\bigO(N^2)$        &   0   & $n+1$ & \multirow{2}{*}{$\bigO(N)$} & \multirow{2}{*}{$n+1$} \\
 & Khan~\cite{Khan2019}    & $\bigO(N \log_2 N)$ & $n-2$ & $2n-1$ \\[5pt]
IFRQI & Khan~\cite{Khan2019} & $\bigO(pN \log_2 N)$        &   $n-2$   & $2n+p-2$ & $\bigO(pN)$ & $n+p$ \\[5pt]
NEQR & Zhang et al.~\cite{Zhang2013neqr} & \multirow{2}{*}{$\bigO(\ell N \log_2 N)$}        &   \multirow{2}{*}{$n-2$}   & \multirow{2}{*}{$2n+\ell-2$} & \multirow{2}{*}{$\bigO(\ell N)$} & \multirow{2}{*}{$n+\ell$} \\
INEQR & Jiang et al.~\cite{Jiang2015a} & & & & \\[5pt]
MCRQI & Sun et al.~\cite{Sun2011} & $\bigO(3N^2)$ & 0 & $n+3$ & $\bigO(3N)$ & $n+3$ \\[5pt]
NCQI & Sang et al.~\cite{Sang2016} & $\bigO(3\ell N \log_2N)$ & $n-2$ & $2n+3\ell-2$ & $\bigO(3\ell N)$ & $n+3\ell$ \\[5pt]
INCQI & Su et al.~\cite{Su2021} & $\bigO(4\ell N \log_2N)$ & $n-2$ & $2n+4\ell-2$ & $\bigO(4\ell N)$ & $n+4\ell$ \\
\bottomrule
\end{tabularx}
\caption{Summary of gate complexities and qubit count for preparing the different QIR states covered in this paper and QPIXL for an image with $N = 2^n$ pixels.
For the IFRQI state, the bit depth is given by $2p$ and for the (I)NEQR, MCRQI, and (I)NCQI states the bit depth is given by $\ell$.}
\label{tab:complexities-all}
\end{table}

\subsection{Further extensions}

We remark that multiple extensions and combinations of the ideas presented in this section are possible.
For example, where MCRQI is a color version of FRQI and (I)NCQI
is a color version of NEQR, we can similarly define a color version
of IFRQI.
We can also adapt IFRQI to group an arbitrary number of bits 
instead of the two bit pairing from \Cref{def:IFRQI}.
This reduces the required number of qubits and gates at the cost of quantum
states that are less distinguishable and thus require more measurements.
It is even possible to use different QPIXL mappings for different RGB color channels. For example, we can use an FRQI mapping for the red channel, an IFRQI mapping for the green channel, and an NEQR mapping for the blue channel.

Finally, although we have presented this discussion for image data in an RGB($\alpha$) space, as in the work by Sun et al.~\cite{Sun2013}, our approach can be readily adapted to different color spaces and even multi-spectral or hyper-spectral
data.
In fact, different scientific applications frequently use images in different color spaces depending on the type of analysis needed.
For example, the Y'CbCr space is known for its applicability to image compression.
The I1I2I3 was created targeting specifically image segmentation.
The HED space is advantageous in the medical field for the analysis of specific tissues.

Similarly, multi-spectral and hyper-spectral data are used in areas such as geosciences and biology, for example, where experts acquire different satellite images and mass spectrometry images respectively.
In all these cases, our general definition of quantum pixel representations can be directly applied. 


\section{\secexamples}
\label{sec:examples}

This section describes a series of experiments that illustrate our proposed tools implemented in \texttt{QPIXL++}~\cite{qpixlpp}.
The current version of \texttt{QPIXL++} supports the FRQI mapping from \Cref{def:FRQI}
for grayscale image data of arbitrary dimensions.

Our first experiment replicates a result from Le et al.~\cite{Le2011b} with
our $\UR$ circuit and compares the gate complexities.
In this test, we consider 10 images with an $8 \times 4$ resolution containing representations of the digits 0 to 9 as shown in \Cref{fig:digits}. These binary images only
contain black and white pixels. We require 5 qubits to encode the pixel location as we have 32 pixels in total.

\begin{figure}[hbtp]
\centering\small%
\begin{tabularx}{\textwidth}{ll|CCCCCCCCCC}
\multicolumn{2}{c}{} 
&\begin{tikzpicture}[scale=0.25]
\fill[black] (0,0) rectangle (1,8);
\fill[black] (1,0) rectangle (3,1);
\fill[black] (3,0) rectangle (4,8);
\fill[black] (1,7) rectangle (3,8);
\draw [step=1,gray] (0,0) grid (4,8);
\end{tikzpicture}
&\begin{tikzpicture}[scale=0.25]
\fill[black] (3,0) rectangle (4,8);
\draw[step=1,gray] (0,0) grid (4,8);
\end{tikzpicture}
&\begin{tikzpicture}[scale=0.25]
\fill[black] (0,0) rectangle (4,1);
\fill[black] (0,4) rectangle (4,5);
\fill[black] (0,7) rectangle (4,8);
\fill[black] (0,1) rectangle (1,4);
\fill[black] (3,5) rectangle (4,7);
\draw [step=1,gray] (0,0) grid (4,8);
\end{tikzpicture}
&\begin{tikzpicture}[scale=0.25]
\fill[black] (3,0) rectangle (4,8);
\fill[black] (0,7) rectangle (3,8);
\fill[black] (0,4) rectangle (3,5);
\fill[black] (0,0) rectangle (3,1);
\draw [step=1,gray] (0,0) grid (4,8);
\end{tikzpicture}
&\begin{tikzpicture}[scale=0.25]
\fill[black] (3,0) rectangle (4,8);
\fill[black] (0,4) rectangle (3,5);
\fill[black] (0,5) rectangle (1,8);
\draw [step=1,gray] (0,0) grid (4,8);
\end{tikzpicture}
&\begin{tikzpicture}[scale=0.25]
\fill[black] (0,0) rectangle (4,1);
\fill[black] (0,4) rectangle (4,5);
\fill[black] (0,7) rectangle (4,8);
\fill[black] (3,1) rectangle (4,4);
\fill[black] (0,5) rectangle (1,7);
\draw [step=1,gray] (0,0) grid (4,8);
\end{tikzpicture}
&\begin{tikzpicture}[scale=0.25]
\fill[black] (0,0) rectangle (1,8);
\fill[black] (1,0) rectangle (3,1);
\fill[black] (3,0) rectangle (4,5);
\fill[black] (1,4) rectangle (3,5);
\draw [step=1,gray] (0,0) grid (4,8);
\end{tikzpicture}
&\begin{tikzpicture}[scale=0.25]
\fill[black] (0,7) rectangle (3,8);
\fill[black] (3,0) rectangle (4,8);
\draw[step=1,gray] (0,0) grid (4,8);
\end{tikzpicture}
&\begin{tikzpicture}[scale=0.25]
\fill[black] (0,0) rectangle (1,8);
\fill[black] (1,0) rectangle (3,1);
\fill[black] (3,0) rectangle (4,8);
\fill[black] (1,7) rectangle (3,8);
\fill[black] (1,4) rectangle (3,5);
\draw [step=1,gray] (0,0) grid (4,8);
\end{tikzpicture}
&\begin{tikzpicture}[scale=0.25]
\fill[black] (0,4) rectangle (1,8);
\fill[black] (0,0) rectangle (3,1);
\fill[black] (3,0) rectangle (4,8);
\fill[black] (1,7) rectangle (3,8);
\fill[black] (1,4) rectangle (3,5);
\draw [step=1,gray] (0,0) grid (4,8);
\end{tikzpicture} \\[5pt]
\multirow{2}{*}{Le et al.~\cite{Le2011,Le2011b}}
 & $\RY$ & 930 & 279 & 930 & 837 & 837 & 930 & 837 & 744 & 930 & 930 \\
 & $\CX$ & 920 & 276 & 920 & 828 & 828 & 920 & 828 & 736 & 920 & 920 \\[5pt]
\multirow{2}{*}{QPIXL}
 & $\RY$ & 8  &  4 &  32 &  32 &  20 & 32 & 32 & 32 & 16 & 11 \\
 & $\CX$ & 20 &  4 &  32 &  32 &  20 & 32 & 32 & 32 & 20 & 16 \\
\midrule
\multirow{2}{*}{Reduction [\%]}
 & $\RY$ & 99.1\% & 98.6\% & 96.5\% & 96.2\% & 97.6\%
         & 96.5\% & 96.2\% & 95.6\% & 98.3\% & 98.8\% \\
 & $\CX$ & 97.8\% & 98.5\% & 96.5\% & 96.1\% & 97.6\%
         & 96.5\% & 96.1\% & 95.6\% & 97.8\% & 98.3\% \\
\end{tabularx}
\caption{$8 \times 4$ image data containing digits 0 to 9,
experiment replicated from Le et al.~\cite{Le2011b}.
Gate complexities for the 6-qubit $\UR$ circuits that prepare an exact representation of the
image data. The last two rows provide the reduction in gate count for our method compared to Le et al.~\cite{Le2011,Le2011b}.
All circuits contain 5 Hadamard gates to create an equal superposition over the first register.}
\label{fig:digits}
\end{figure}

The method of Le et al.~\cite{Le2011} requires one
$C^5(R_y)$ gate for every pixel, bringing the total
up to $32$ $C^5(R_y)$ gates.
Every $C^5(R_y)$ gate is further decomposed into 93 $\RY$
and 92 $\CX$ gates.
The experiment described by Le et al.~\cite{Le2011b} reduces the number of
$C^5(\RY)$ gates through a compression algorithm that groups
pixels with the same grayscale value.
This method is effective for the binary data in \Cref{fig:digits}
as they report lossless compression ratios between $68.75\%$ and $90.63\%$.
\Cref{fig:digits} compares the number of 1-qubit $\RY$ and $\CX$ gates for our method with the results from Le et al.~\cite{Le2011b}.
We ran our compression algorithm with a compression level of 0\% to the $\UR$ circuit.
Thus only coefficients in $\hat \bftheta$ that are exactly 0 are removed, which means that
our circuits are exact.

\Cref{fig:digits} shows that our method always provides more than 95\%
reduction in gate count compared to the method from Le et al.~\cite{Le2011,Le2011b} for this example.
The advantage of our method becomes even more outspoken for larger images due to the quadratic improvement.

The next example we present concerns an image taken from the MNIST database~\cite{MNIST} of handwritten digits.
The image of the digit ``3'' has a resolution of $28 \times 28$ pixels that is zero padded to an image with 1024 pixels in \texttt{QPIXL++} which means that roughly 75\% of
the coefficients are used for the actual image data.

\begin{figure}[hbtp]
\centering\small%
\begin{tabularx}{\textwidth}{ll|CCCCC}
\multicolumn{2}{c}{} & 0\% & 30\% & 60\% & 75\% & 90\% \\[2pt]
\multicolumn{2}{c}{} 
 & \includegraphics[scale=2]{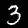}
 & \includegraphics[scale=2]{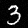}
 & \includegraphics[scale=2]{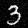}
 & \includegraphics[scale=2]{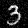}
 & \includegraphics[scale=2]{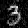} \\[5pt]
\multirow{2}{*}{Gate count}
 & $\RY$ & 1024 &  717 &  410 &  256 &  103 \\
 & $\CX$ & 1024 &  914 &  666 &  494 &  224 \\
\midrule
\multirow{2}{*}{Reduction [\%]}
 & $\RY$ & 0.0\% & 30.0\% & 60.0\% & 75.0\% & 90.0\% \\
 & $\CX$ & 0.0\% & 10.7\% & 35.0\% & 51.8\% & 78.1\% \\
\end{tabularx}
\caption{$28 \times 28$ image data from the MNIST database simulated
with~\texttt{QPIXL++} at various compression levels and corresponding
gate counts of the 11-qubit $\UR$ circuit.
The final two rows list the reduction in $\RY$ and $\CX$ gates compared to the uncompressed circuits.}
\label{fig:MNIST}
\end{figure}

\Cref{fig:MNIST} shows the images that are simulated with \texttt{QPIXL++} at 5 different compression levels.
There are no visual artifacts at 30\% compression and also the image at 60\%
compression is close to the original quality. 
The image with a 75\% compression ratio has more visual artifacts but is still clearly recognizable, while at 90\% compression the quality begins to drop significantly.
The corresponding gate complexities for the $\UR$ circuits are also listed in~\Cref{fig:MNIST}, all circuits contain 10 Hadamard gates to create the superposition in the first register.
We observe that the reduction in $\RY$ gates is in perfect agreement with the compression ratio, but that there is generally a smaller reduction in $\CX$ gates. 
This is in line with the expectations for our proposed compression algorithm described in~\refcompr: not all $\CX$ gates along a sequence of removable $\RY$ gates will cancel out.
This experiment in particular clearly identifies a potential application of our QIR with compression to classification algorithms based on machine learning in quantum computers.

\begin{figure}[hbtp]
\centering\small%
\begin{tabularx}{\textwidth}{ll|CCC}
\multicolumn{2}{c}{} & 0\% & 50\% & 75\% \\[2pt]
\multicolumn{2}{c}{} 
 & \includegraphics[scale=0.4]{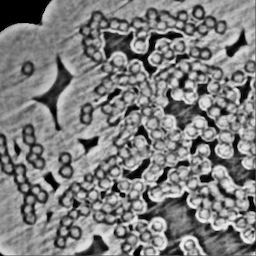}
 & \includegraphics[scale=0.4]{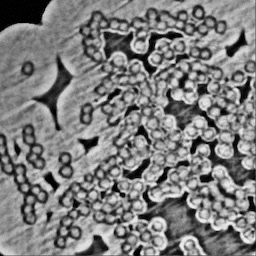}
 & \includegraphics[scale=0.4]{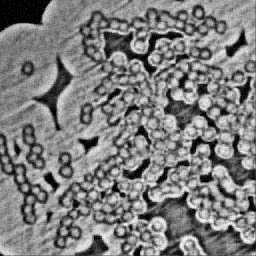} \\[5pt]
\multirow{2}{*}{Gate count}
 & $\RY$ & 65,535 & 32,768 & 16,384 \\
 & $\CX$ & 65,536 & 44,694 & 25,040 \\
\midrule
\multirow{2}{*}{Reduction [\%]}
 & $\RY$ & 0.0\% & 50.0\% & 75.0\% \\
 & $\CX$ & 0.0\% & 31.8\% & 61.8\% \\
\multicolumn{5}{c}{} \\[10pt]
\multicolumn{2}{c}{} & 90\% & 95\% & 99\% \\[2pt]
\multicolumn{2}{c}{} 
 & \includegraphics[scale=0.4]{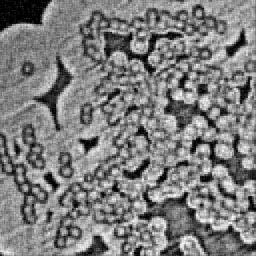}
 & \includegraphics[scale=0.4]{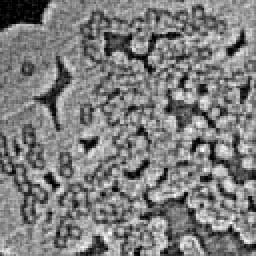}
 & \includegraphics[scale=0.4]{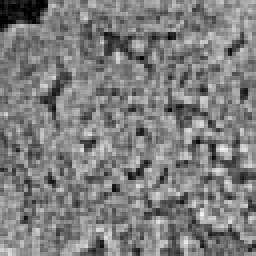} \\[5pt]
\multirow{2}{*}{Gate count}
 & $\RY$ &  6,554 & 3,277 &   656 \\
 & $\CX$ & 11,090 & 5,852 & 1,402 \\
\midrule
\multirow{2}{*}{Reduction [\%]}
 & $\RY$ & 90.0\% & 95.0\% & 99.0\% \\
 & $\CX$ & 83.1\% & 91.1\% & 97.9\% \\
\end{tabularx}
\caption{$256 \times 256$ image data from of a ceramic matrix composite sample acquired using microCT simulated with \texttt{QPIXL++} at various compression levels and corresponding
gate counts of the 17-qubit $\UR$ circuit.
The final two rows list the reduction in $\RY$ and $\CX$ gates compared to the uncompressed circuits.}
\label{fig:ex3}
\end{figure}

Our final example image stems from scientific data.
This is a $256 \times 256$ pixels region from a cross-section of a ceramic matrix composite (fiber reinforced polymer)~\cite{Bale2013RealtimeQI} imaged with X-ray micro computed tomography (microCT) at the LBNL ALS beamline 8.3.2.
This type of image is frequently acquired by material scientists to study the development of material deformation under stress.
Consequently, image analysis algorithms to detect the circular patterns present in the image for example (cross-sections of fibers) become extremely important.
As the dimensions of this grayscale image are already a power of 2, it does not need to be zero-padded. 
It contains both large scale structure and fine scale details.
We require 16 qubits to encode the pixel locations and 1 for the grayscale intensities
such that the $\UFRQI$ circuit has a total of 17 qubits.
The uncompressed $\UR$ circuit contains $2^{16}$ or 65,536 $\CX$ and $\RY$ gates.
We ran our compression algorithm on the data and the results 
are summarized in \Cref{fig:ex3}.

As can be observed, the compression algorithm is very
effective for this image. Up to 75\% compression can be achieved while still
maintaining both the large scale structure and the finer details.
The large scale structure is still preserved at 95\% compression, but the
acuteness in the finer details is lost at this compression level.
It is only at 99\% compression that the image becomes completely 
dominated by compression artifacts.
It becomes clear from this last example that our compression approach becomes extremely interesting when analyzing scientific data: (1) the amount of data to be processed is reduced, and (2) the approach maintains details in the image necessary for further analysis, such as feature extraction for example.

\begin{figure}[hbtp]
\centering
\figname{scaling}%
\begin{tikzpicture}
\begin{axis}[
  xmin=1,xmax=34,%
  ymin=1e-8,ymax=3e2,%
  ymode=log,%
  xlabel={number of pixels $\log_2(N)$},%
  ylabel={time [s]},%
  legend style={draw=none,fill=none,row sep=-1pt},%
  legend pos=north west,%
]
\addplot[myColOne,mark=]%
   table[x index=0,y index=1] {\datfile{fwht}};
\addplot[myColTwo,mark=]%
   table[x index=0,y index=1] {\datfile{gray}};
\legend{sFWHT,
        Gray permutation};
\addplot[no marks,gray,densely dotted] plot coordinates {(3,1e-7) (30,134)};
\draw[gray] (25,30) node[below left,rotate=40] {\footnotesize$\bigO(N \log_2(N))$};
\end{axis}
\end{tikzpicture}%
\caption{\label{fig:scaling}%
Scaling for scaled fast Walsh-Hadamard transform (sFWHT) and in-place Gray permutation
with \texttt{QPIXL++}.}
\end{figure}
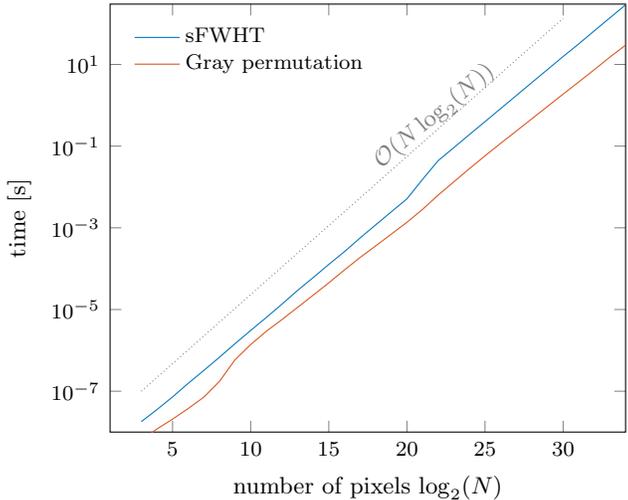

To conclude the experiments section, we present benchmark data for solving the
linear system~\eqref{eq:ls} with the matrix-free methods that are implemented in \texttt{QPIXL++}.
These timing results are obtained on an AMD Ryzen Threadripper 3990X 64-Core Processor
@ 2.9 GHz with 256 GB RAM.
The results are shown in \Cref{fig:scaling} for randomly
generated image data ranging from $2^3$ pixels up to $2^{34}$ pixels. The latter corresponds to the equivalent of an image with a resolution of more than 17 gigapixels,
a 4K video fragment with 2070 frames, or a 1080p video fragment with 8285 frames.

Computing the coefficients for the uncompressed FRQI
circuit with $2^{34}$ pixels requires just over 5 minutes.
The bottleneck is the memory required to store the image data. 
This experiment shows that our method easily scales to high resolution image and video data.


\section{\secconclude}
\label{sec:conclude}

We have introduced an overarching framework for quantum pixel representations and
showed how previously introduced image representations can be incorporated in the QPIXL framework.
Among these methods are (I)FRQI, (I)NEQR, MCRQI, and (I)NCQI.
We have proposed a novel circuit synthesis technique for preparing the quantum pixel representations
on a quantum computer.
This technique makes use of uniformly controlled $\RY$ rotations
and significantly reduces the gate complexity for all aforementioned methods.
Hence, the obtained circuits only require $\RY$ and $\CX$ gates which makes them feasible for the NISQ era.
Our method requires the solution of a particular linear system which can be solved classically in
$\bigO(N \log N)$ time with a matrix-free approach.
Furthermore, it allows for an efficient image compression algorithm 
that works on the transformed image data.
Our experiments show that this compression approach is very effective for the FRQI mapping
and can further reduce the
number of gates by as much as 90\% while still retaining the most prominent features of the image
in the FRQI state.
We repeatedly show how our method can have great impact on the analysis of scientific data and for quantum machine learning applications in the future.
We have implemented and tested our algorithms in a publicly available software package
\texttt{QPIXL++}~\cite{qpixlpp} which supports QASM output.
Benchmark timings show that \texttt{QPIXL++} has excellent scaling properties and can
handle high resolution image and video data.

\bibliographystyle{abbrvurl}
\bibliography{references}

\end{document}